# Controlling of ZnO nanostructures by solute concentrationand its effect on growth, structural and optical properties

Yogendra Kumar[1,†], Amit Kumar Rana[2,†], Prateek Bhojane[1], Manojit Pusty[1], Vivas Bagwe[3], Somaditya Sen[1,2] and Parasharam M. Shirage[1,2,*]

[1]Centre for Materials Science and Engineering, Indian Institute of Technology Indore, Khandwa Road, Simrol Campus, INDORE - 452020. INDIA.

[2]Department of Physics, Indian Institute of Technology Indore, Khandwa Road, Simrol Campus, INDORE - 452020. INDIA.

[3]DCMPMS, Tata Institute of Fundamental Research, Colaba, Mumbai-400005. INDIA.

*Author for correspondence E-mail: paras.shirage@gmail.com, pmshirage@iiti.ac.in

[†] Share equal contributions

**Abstract**

ZnO nanostructured films were prepared by chemical bath deposition method on glass substrates without any assistance of either microwave or high pressure autoclaves. The effect of solute concentration on the pure wurtzite ZnO nanostructure morphologies is studied. The controlling of the solute concentration help to control nano-structure in the form of nano-needles, and -rods. XRD diffraction studies revealed highly *c*-axis oriented thin films. SEM confirms the modifications in the nanostructure dependent on the concentration. TEM results shows the single crystalline electron diffraction pattern indicating high quality nano-material. UV-Vis results shows the variation in band gap from 3.20 eV to 3.14 eV with increasing the concentration as nano-structures modifies from needle to rod like structure. Photoluminescence (PL) data indicates existence of defect in the nanomaterials and emitting light in the yellow-green region, with broad UV and visible spectra. The sharp and strong peak in Raman spectroscopy is observed at ~438 cm$^{-1}$, assigned to the $E_2^{high}$ optical mode of the ZnO, characteristic peak for the highly-crystalline wurtzite hexagonal phase. Solute concentration significantly affects the formation of the defect states in the nanostructured films as result it alters the structural and optical properties. Current–Voltage characteristics alters with measurement environment, indicating the potential sensor applications.

# 1. Introduction

Over the past few decades, the extensive studies on zinc oxide (ZnO) has been carried out for its applications in different electro-optical industries and technologies, such as spintronics devices [1], light emitting [2] and laser diodes [3] surface acoustic wave devices [4] and transparent electrodes for photovoltaic (solar) cells [5] and field emission displays [6]. For the progress of ZnO-based optoelectronic devices, it is indispensable to grow high-quality ZnO thin films with both *n*-type and *p*-type electrical conductions. As grown ZnO is the II-VI transparent *n*-type metal oxide semiconductor with a wide direct band gap (3.37 eV) and large exciton binding energy of 60 meV at room temperature[7,8] which turn ZnO into an attractive material for an extraordinarily wide spread range of potential opto-electronic application such as UV laser[9], thin film transistors[10], UV detectors[7], piezoelectric devices[11], gas sensors[12], also it has biomedical applications such as antibacterial and antifungal agent[13].Moreover to UV excitonic emission peak, ZnO usually shows visible luminescence at varied emission wavelengths owing to intrinsic or extrinsic defects[14].But for these applications and thoughtfulgeneratingwell-ordered defects, it is crucial to control shape, size and crystallinity of ZnO nanostructures,since physical and chemical properties are meticulously dependenton them. In the past few years researchersare flourished to synthesizediverse type of ZnO nanostructure like nano-plates[15], nano-tube[16], nano-flowers[17], nano-rods and nano-belts [18, 19].Among these nanostructures, preferentially aligned vertical and dense nano-rods on a desired substrate potentially deliver thestriking solution to achieveexcellent properties such as energy harvesting devices, field emission display, and light emitting diodes, *etc*.The use of nanoscale devices and systemsare desirable as scientific society facing a wide range of challenges to fulfilltheobligation. There are several method for getting controlled ZnO nanostructure, such as chemical bath deposition[15], chemical vapor deposition[20], electrodeposition[21], electrospinning [22], precipitation method[23],laser assisted flow deposition[24] and hydrothermal techniques[25-28].Among all these aforementioned different methods, chemical bath deposition method is simple and economical, where the external condition such as pH value, reaction temperature, reaction time, solution concentration *etc*. are the only critical parameters that will affect the reaction rate and then morphologies of ZnO. In the present investigation we have utilized the chemical bath deposition method to grow highly oriented ZnO films and studied the effect of solute concentration on structural, optical and electricalproperties. By controlling the solute concentration the morphologies has been controlled in the form of needles and rods and the characterization has been done by different techniques has been discussed in this report. Also we explore the possibility of using this material in sensing application by measuring current-voltage characteristics in different environment like air, ethanol and acetone.

## 2. Experimental technique
## Synthesis Process

Concentration dependent ZnO nanostructures films on glass substrates were synthesized by a simple and economical chemical bath deposition process in desired scale in different morphologies like, needles and rods. An aqueous solutions of $(Zn(NO_3)_2 \cdot 6H_2O)$ salts ( 50 mM -250 mM at the interval of 50 mM, Alfa Aesar Chemicals) by adding the appropriate quantity of double distilled (DD) water were used as starting materials. After complete dissolution of the solute in DD water, ammonia was added drop wise and maintained p(H)~12. The glass substrates were used as the substrates as they are cheap and suitable for many optical devices. Well cleaned glass substrates are kept vertically in the reaction solution. To understand the growth mechanism more clearly, reaction temperature (120 °C) and time (~90 min) were kept constant. After completion of the reaction time, the coated films were taken out from the solution and rinsed carefully in distilled water several times to remove excess ammonia from the substrate and annealed in air at 300°C for 2 hrs. to get pure phase ZnO nanostructure.

## Experimental details

The phase formation and surface morphology of ZnO films were characterized by a X-ray diffraction ( XRD, Bruker D8 Advance X-ray diffractometer) with Cu-$K_\alpha$ radiation ($\lambda$= 1.54 Å) and Field Emission Scanning Electron Microscope ( FESEM, Supra 55 Zeiss), respectively. Additional structure analysis of distinct nanostructures was carried out using transmission electron microscope (TEM). TEM images were taken by HRTEM, JEOL JEM 2100 and JEOL 3010 with a UHR polepiece with selective electronic area diffraction (SEAD). Optical band gaps were determined with the Diffuse Reflectance Spectrometer (Agilent Cary-60 UV-Vis). RT photoluminescence spectra's of ZnO films were conducted using a Dongwoo Optron spectrophotometer. The excitation wavelength was fixed at 325 nm and emission spectra was scanned from 350 to 850 nm. Micro Raman scattering of the samples were recorded by using Labram-HR 800 spectrometer equipped with excitation radiation with 488 nm wavelength by Argon ion laser at a spectral resolution of about 1 cm$^{-1}$. The *I–V* characteristics of ZnO in presence of different atmosphere were measured by usingprogrammable Keithley source meter 2401.

## 3. Result and Discussion

XRD measurements wereaccomplished to examine the crystal structure and phase purity of ZnO nano-needle and -rods arrays. Fig.1. shows the XRD patterns of the ZnO nanostructures on glass substrates with different concentrations, (*a*) 50mM, (*b*) 100mM, (*c*) 150mM, (*d*) 200mM,and (*e*) 250mM deposited atconstant temperature and time (120 °Cand 90 min.), respectively.

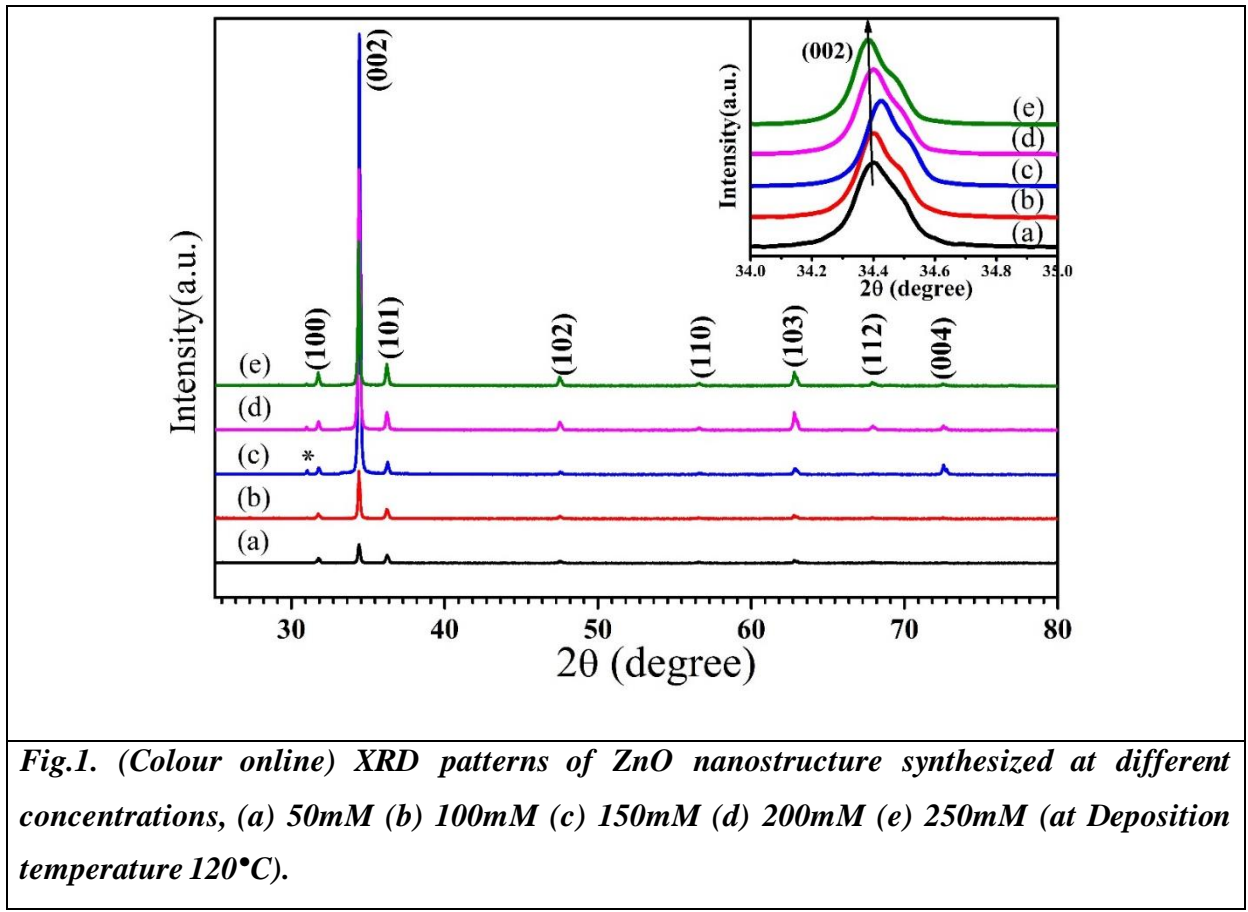

*Fig.1. (Colour online) XRD patterns of ZnO nanostructure synthesized at different concentrations, (a) 50mM (b) 100mM (c) 150mM (d) 200mM (e) 250mM (at Deposition temperature 120°C).*

All the major peaks in the range ($2\theta \sim 25° - 80°$),of all the films can be indexed to the hexagonal phase of wurtzite ZnO (space group *P63mc*), which are rendering to JCPDS card number 80-0075 and suggesting that the films are phase pure. However, a very low intensity peak nearby $2\theta = 31.03°$ is observed for $Zn(OH)_2$. It is evidence from the Fig.1 that all the films show strong directional growth along (002), which is unusual compared to other ZnO nanostructures[29-30].Fig.1 shows the negligible variation in peak position with the variation inconcentration of the ZnO film,this indicates the lattice parameters are insignificantlychanges with changing the concentration of ZnO.

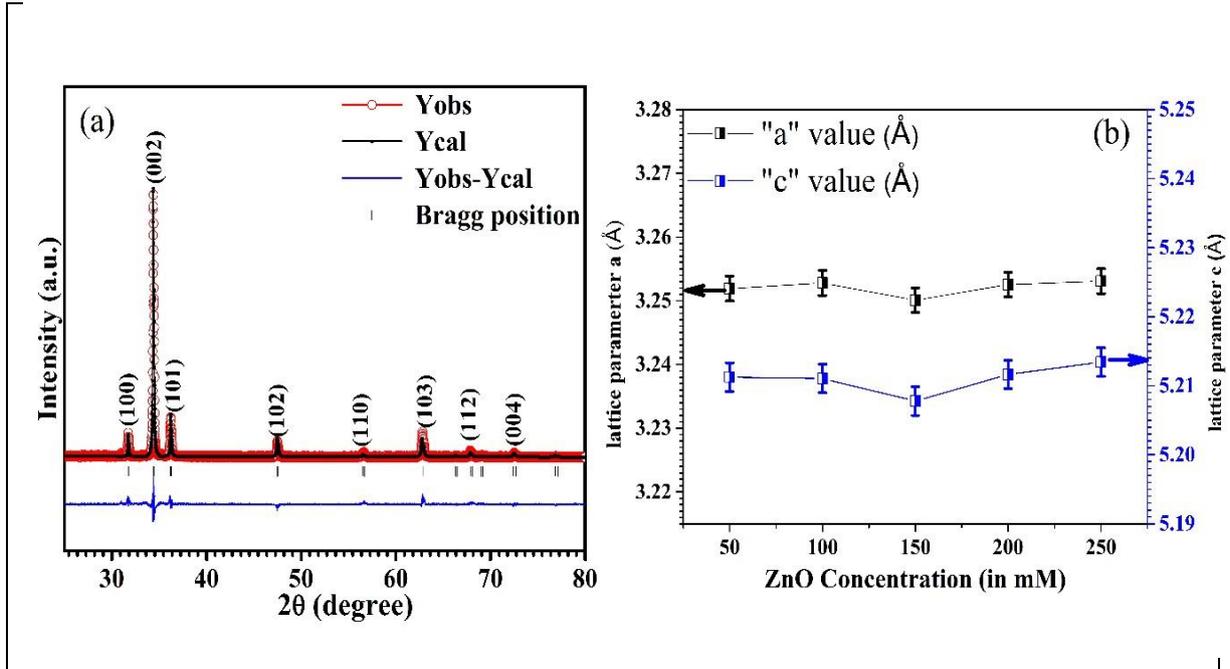

*Fig.2. (Color online) (a) Reitveld refinement of ZnO deposited with concentration of 250 mM. (b) Shows the lattice parameter variation in ZnO with different concentration.*

Reitveld refinement (Fig. 2 (*a*)) was performed using "Fullprof" software and the values of the refined lattice parameters obtained from reitveld fitting are shown in Fig.2 (*b*). Fig.2 (*b*) shows the variation of lattice parameters as a function of concentration. The concentration shows negligibly small effect on lattice parameters. There might be probable reasons for the nonsystematic change in lattice parameter with concentration, it is due to internal compressive micro strain [31]. For Size determination and effect of strain developed in the nano structure grown on the glass substrates were analyzed using Debye Scherre's formula and Williamson-Hall [32] plot using uniform deformation model (UDM) (Table I). Expressed as

$$D = \frac{K\lambda}{\beta_D cos\theta} \quad\quad\quad (1)$$

and

$$\beta_{hkl} = \frac{K\lambda}{Dcos\theta} + 4\varepsilon tan\theta \quad\quad\quad (2)$$

Where D = crystalline size, $K$ = shape factor (0.9), $\lambda$ = wavelength of $Cu_{k\alpha}$ radiation and $\beta_D$ is full width half maxima of a corresponding peak and $\varepsilon$ is induced strain value.

| Sample | Debye Scherre's method | Williamson- Hall method Uniform Deformation Model (UDM) | |
|---|---|---|---|
| | D (nm) | D (nm) | Strain ($\varepsilon$) x $10^{-3}$ |
| 50 mM | 45.715 | 46.321 | 1.92 |
| 100 mM | 46.205 | 46.986 | 2.40 |
| 150 mM | 47.259 | 48.171 | 2.69 |
| 200 mM | 48.361 | 49.356 | 2.80 |
| 250 mM | 49.500 | 50.608 | 2.97 |

**Table I. Shows the value of crystalline size and strain ($\varepsilon$) for the ZnO nano structures grown at different concentrations obtained by Debye Scherre's method and Williamson-Hall method**

All the parameters obtained by using Debye Scherre's and Williamson-Hallformula are listed in Table I. From Table I it can be seen that strain is dependent on the crystallite size of the nanostructures and higher the crystallite size, higher is the strain and vice versa.

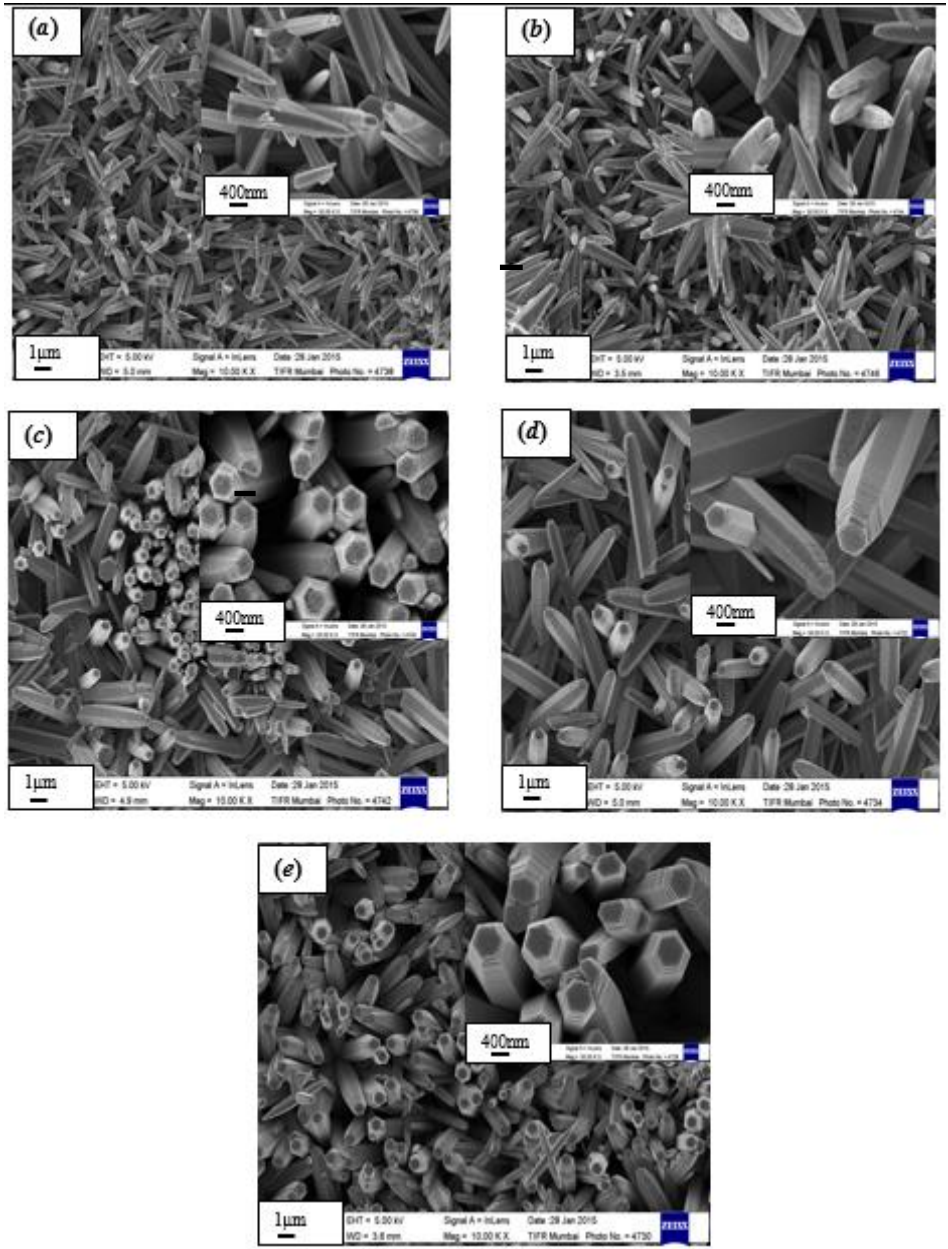

**Fig.3. 10 KX FESEM images of ZnO deposited on glass substrate at different concentration of (a) 50mM (b) 100mM (c) 150mM (d) 200mM and (e) 250mM. (Inset figure shows images at 50 KX).**

Variation of surface morphology with solute concentration for the deposition of ZnO nanostructed films were studied by FESEM and represented in Fig. 3. It shows that films surface consists of nano-needles, tapred nanorods and nanorods like microstructures with average length of 3μm to 7μm and average diameter is 100 nm to 600 nm. All films are composed of nano-needles or nano-rods like structures. With increase in the concentration, the increase in diamter of the nano-needles/rods was found (~150nm, ~200nm, ~450 nm, ~600nm and ~650nm for 50mM, 100mM, 150mM, 200mM and 250mM, respectively). A higher concentration produces a micro-sized diameter with densely packed *c*-axis aligned ZnO rods as shown in fig.3 (*e*).ZnO films deposited at lower concentration (50 mM) shows the tower (needle) shape like nanorods. As thesolute concentration of ZnO increased, due to the high nucleation growth, nano-needles/rods starts the formation of stacking of flate tips i.e.splitting of bigger rods into smaller rods which is very common in ZnO [27]. With further increase in the concentration, the diameter of nanorod increased and forms a flate tip like nanorod with larger diameter. Screw dislocation nucleation mechanism is one of the responsible mechanims of nano-needles and rods as we reported in ref [28,33], as we can rule out the layer by layer deposition mechanism. Fig.4 showsthe schematic of the transformation of nano-needes to nano-rods with variation in concentration during synthesis of ZnO films. It is clear from this schematic and FESEM images lower concentration results in needles and higher concentration to rods.

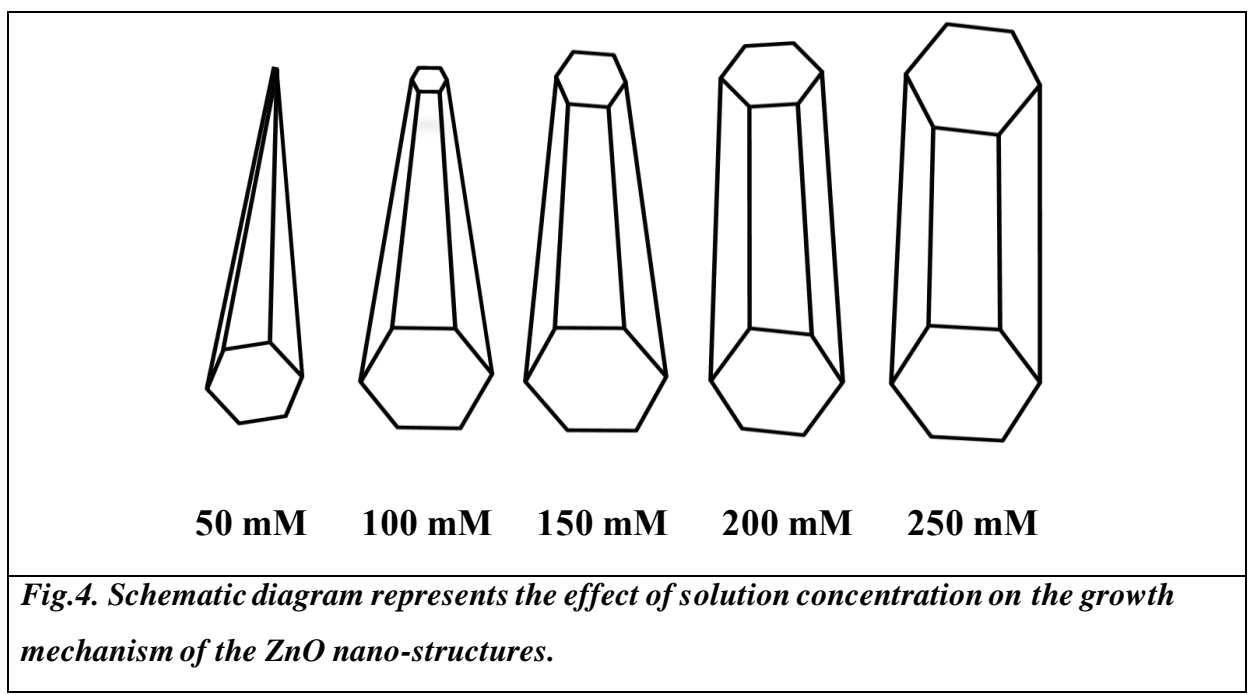

*Fig.4. Schematic diagram represents the effect of solution concentration on the growth mechanism of the ZnO nano-structures.*

The size and morphology of ZnO nanostructure is analyzed by TEM and is represented in fig.5. Fig.5 (*a*) shows the TEM image of one of the ZnO nanorod synthesized (100 mM) by the chemical bath deposition method. It shows ZnO nano rod with diameter around ~200 nm. Also, the corresponding SAED pattern (*fig*.5 (*b*)) confirms that the nanorod has a single-crystalline wurtzite structure growing along the *c*-axis. XRD and TEM confirms that the present employed method results in nano-materials with *c*-axis orientation irrespective of the concentration, which is very useful for the opto-electronics device fabrications.

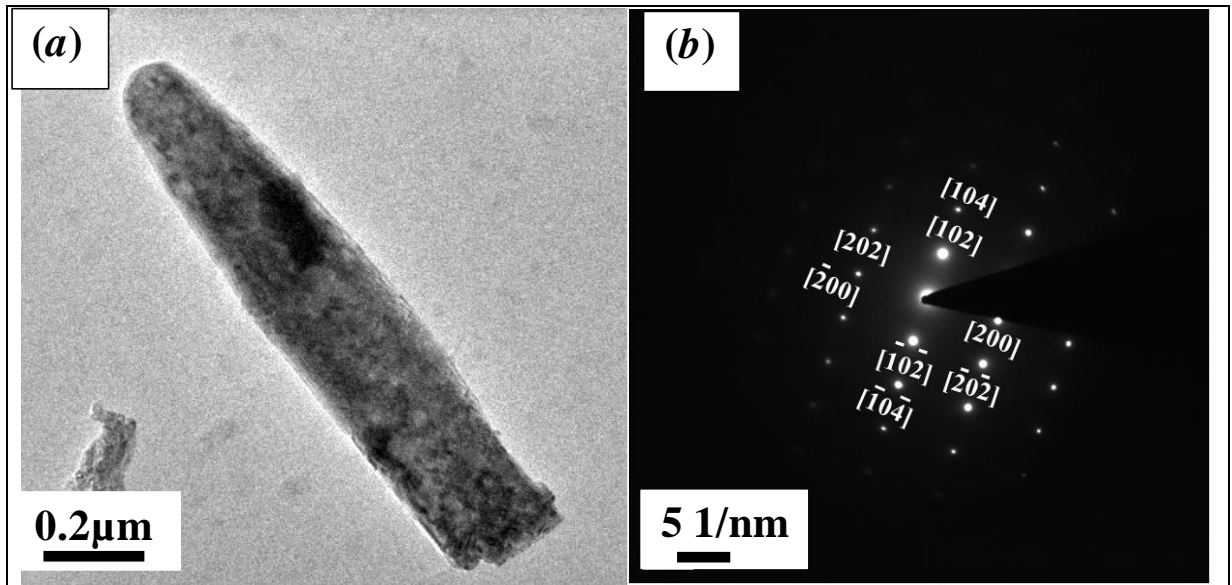

*Fig.5(a) Transmission electron microscopy (TEM) image of a single ZnO nanorod (b) Selected area electronic diffraction(SAED) pattern of thenanorods structure. The planes from which the electron beam was diffracted to generate the diffraction pattern are indexed in the diffractogram.*

Room temperature UV-Vis spectroscopy (in the range of 300 to 800 nm) was used to obtain the intrinsic optical properties (band-gap) of ZnO nano-structures synthesized at different concentrations. Fig. 6 show the UV-visible spectra of ZnO from the range 380 nm to 425 nm. It is noticeable that the band gap is varying from 3.20 eV to 3.14 eV as the concentration increases, which is due to change in morphology of ZnO nanostructure from nano-needles to nano-rods. This result clearly indicates that the optical properties like band-gap of ZnO strongly depends on the morphology of the nano-structures, the band gap was determined from Kubelka- Munk method [34].

$$F(R_\infty)h\nu = A(h\nu - E_g)^n$$

Where, $F(R_\infty)$ is the Kubelka-Munk function, A is a constant, $E_g$ is the band gap value and *n* is an unit less parameter with a value 2 or ½ for indirect or direct band gap semiconductors, respectively [35].

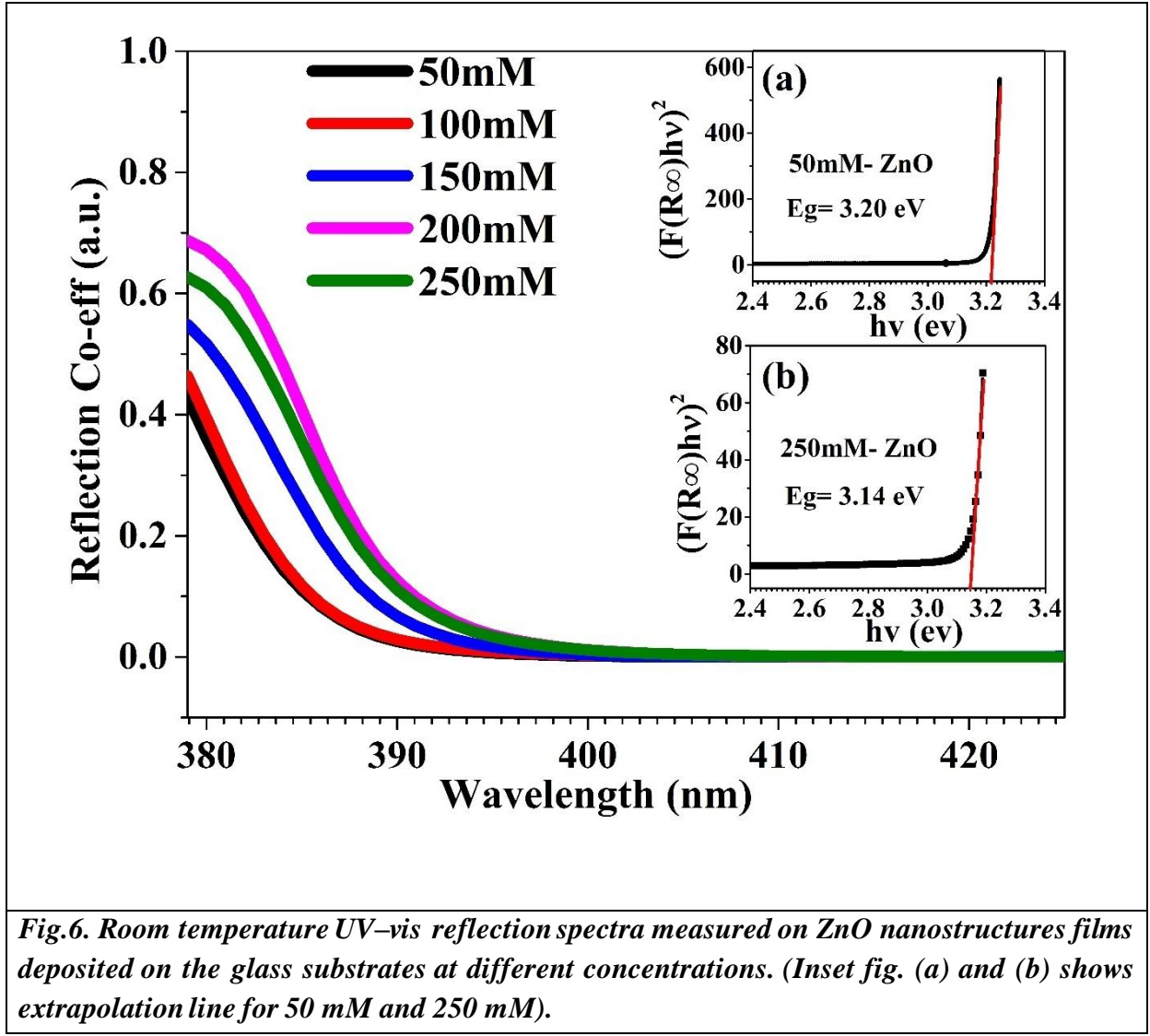

*Fig.6. Room temperature UV–vis reflection spectra measured on ZnO nanostructures films deposited on the glass substrates at different concentrations. (Inset fig. (a) and (b) shows extrapolation line for 50 mM and 250 mM).*

As a typical example the Tauc plot for the ZnO 50mM and 250 mM sample are shown in inset of *fig*.6(*a*) and (*b*). From the inset, it is very clearly shows that the band gap is high (~3.20 eV) in 50mM concentration and smaller (~3.14 eV) at higher concentration of 250mM. Fig.7 shows the variation in band-gap values calculated from Kubelka-Munk method. It is clearly reflected that with increase in solute concentration of ZnO, near band edge (NBE) peak position shifts to lower energy side.

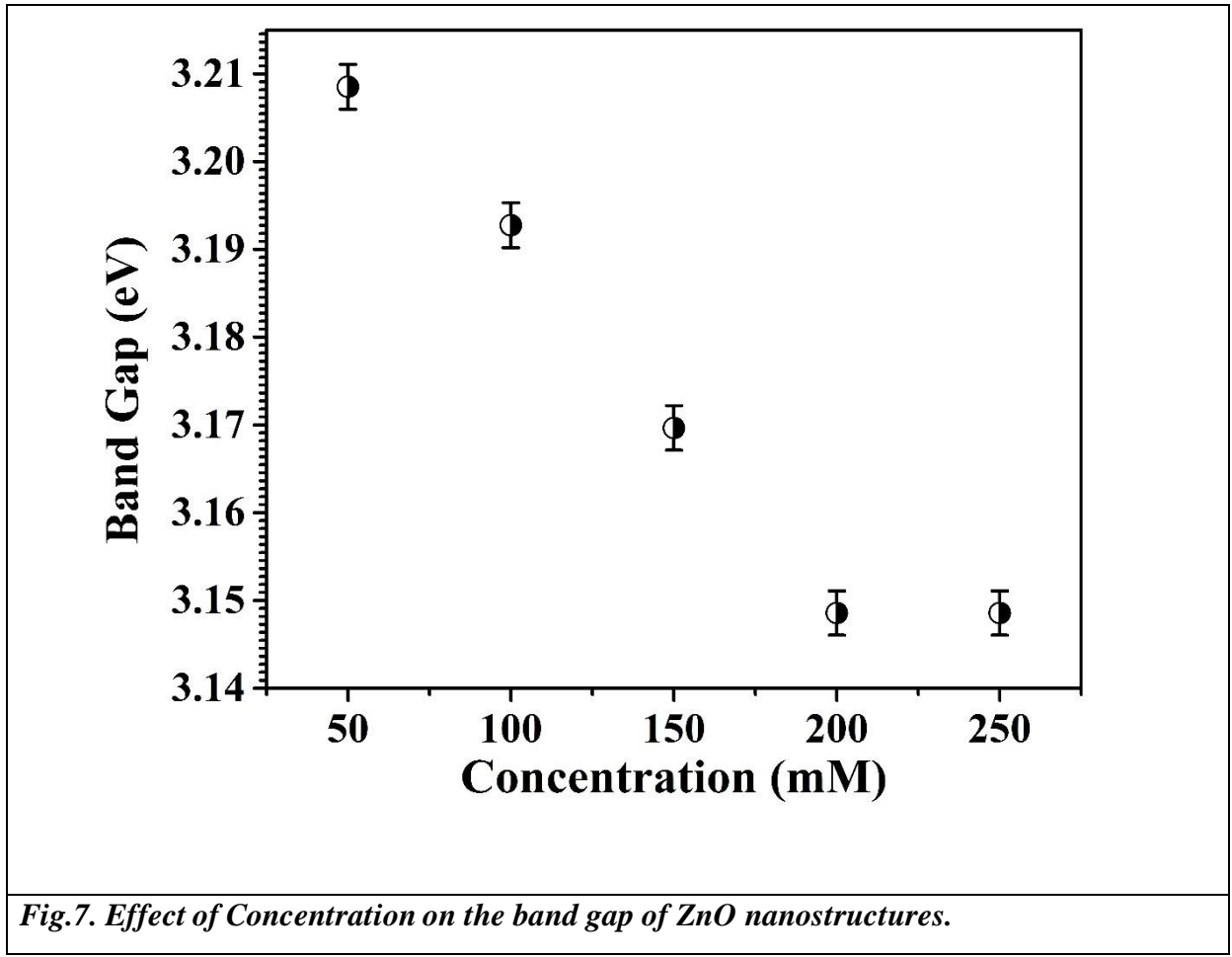

*Fig.7. Effect of Concentration on the band gap of ZnO nanostructures.*

Photoluminescence spectroscopy (PL) is an effective method to investigate the presence of defects in the semiconductors. The photoluminescence (PL) spectra of ZnO nanostructure depended on the preparation condition (deposition time and temperature, concentration *etc.*). Fig.8 displays the room temperature photoluminescence (PL) of ZnO nanostructure synthesized at different concentration with an excited wavelength of 325 nm. It evident from the Fig.8 that all films contain two main peak region, one peak is belongs to UV region and another one is fits to broad green-yellow emission in the visible luminescence (*VL*) region. It indicates that as we increase the concentration the intensity of *VL* region increases then decreases. It indicates that with increasing the concentration, there is an increase in defect levels up to 200mM and then decreases.

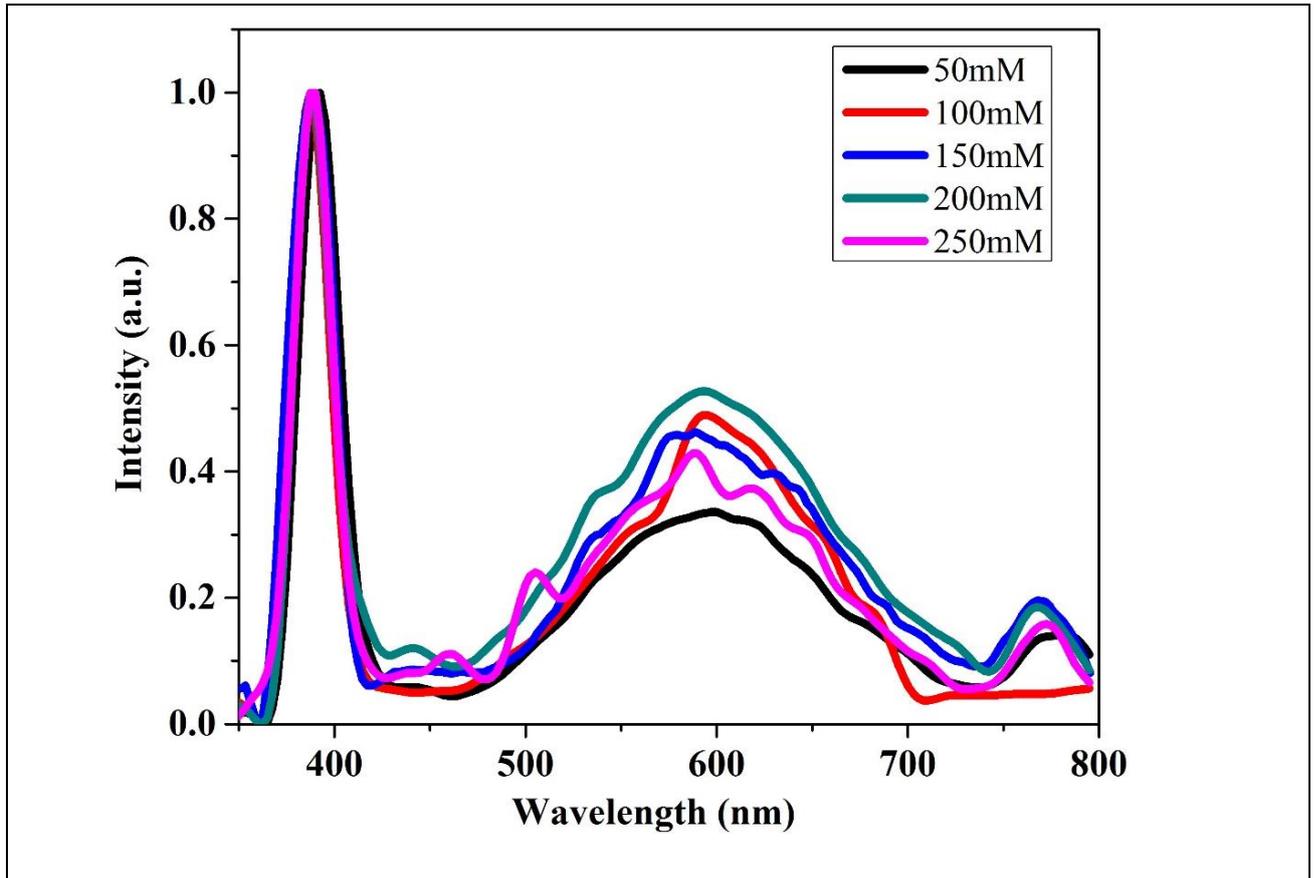

*Fig.8. Room temperature normalized PL spectra of ZnO deposited at different concentrations.*

Fig.9 demonstrates the room temperature photoluminescence (PL) spectra and fitting curves of ZnO (200 mM) spread over a broad range from 350 to 800 nm (3.2 eV to 1.60 eV) measured by exciting with energy of 325 nm. The typical wide emission spectrum of ZnO nanostructures extending from near band emission edge (NBE) to red emission which can be fitted with well resolved 12 peaks. In ZnO nanostructures, existence of defects is very well known such as Zinc vacancy ($V_{Zn}$), Zinc interstitial ($Zn_{in}$), Oxygen interstitials ($O_{in}$), Oxygen vacancy ($V_O$), Oxygen interstitial ($O_{in}$) and Antisite Oxygen have been reported previously. From PL spectrum, a much strong band edge (ultraviolet at 387 nm) is observed resulting from the near band edge emission of the wide band gap ZnO, while in addition to UV excitonic emission peak, ZnO commonly exhibits visible luminescence at different emission wavelengths due to intrinsic or extrinsic defects[27]. We observed violet emission at 401nm, which is believed that the Zinc vacancies ($V_{Zn}$) is responsible [36]. There is blue emission at 440 nm and 489 nm, which is coming from the transition between conduction band and Zinc vacancy or Zinc interstitial to Zinc vacancy. But the origin of these emission is not very clear.

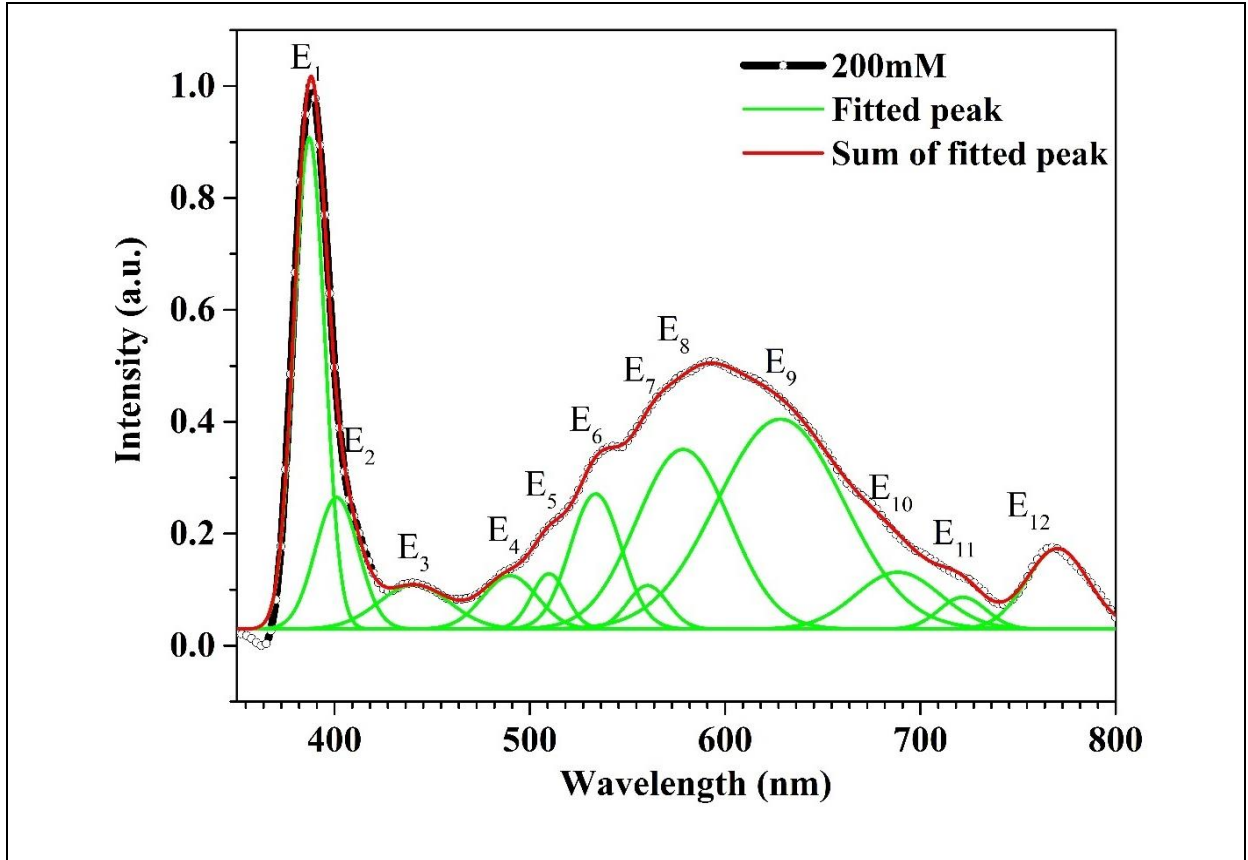

*Fig.9. (Color online) Room temperature PL spectra (Black dots),Gaussian fitting curves(Green) and summation of Gaussian Creve (red line) of ZnO film grown at 200mM concentration.*

We also see another controversial issue is related to green emission, whose origin is unknown, we observed at 509nm, 533nm, and 560nm. These emission is either related to oxygen deficiency, i.e. from conduction band to oxygen vacancy or Zinc vacancy that is not distinguishable. While yellow emission at ~578 nm weak and wider, is associated with excess of oxygen [37] and imminent from the transition from the conduction band to oxygen interstitials. And finally small red emission around 628nm, 688nm, 721nm and 769nm is observed. The origin of red emission is divided into two parts, first is red emission range from 620nm to 690nm can be attribute oxygen interstitial and while in the range of 690nm to 750nm can be attributed to oxygen vacancy. In our case, the progressive change in yellow light emission intensity relative to UV emission suggests that there is a greater fraction of oxygen interstitial in these nanorods. The intensity of this peak varies with the crystallinity of ZnO nanostructure as predicted. Hence, the PL of the ZnO microstructures can be prudentlyemployed by changing their morphologies. All the the peaks appeared and possible defects in PL are summarized in Table 2. These emission peak in the visible region enable thin film for the application as a laser diodes and light emitting diode. PL spectra are strongly dependent on the crystalline nature of the films

**Table II.** *PL peak positions and respective defects assigned in ZnO nanostructures.*

| Peak | Wave length (nm) | Proposed deep level transition (color emission) |
|---|---|---|
| E1 | 387 | Near band emission [27] |
| E2 | 401 | Related to $V_{Zn}$(violet emission)[36] |
| E3, E4 | 440, 489 | CB to $V_{Zn}$ or $Zn_{in}$ to $V_{Zn}$ ( Blue emission) [21,37] |
| E5, E6, E7 | 509, 533, 560 | CB to $V_O$ or to $V_{Zn}$(Green emission )[38-39] |
| E8 | 578 | CB to $O_i$ (Yellow emission) [40-42] |
| E9, E10, E11, E12 | 628, 688, 721, 769 | due to lattice disorder along c-axis, ($Zn_i$ defect)(Red emission) [43] |

The wurtzite-type structure of ZnO indicates a basic unit of four atoms per unit cell with space group $C_{6v}^4$. Due to the four number atoms in the unit cell, ZnO has total 12 phonon mode, with 3 acoustic modes (one LA, two TA) and 9 optical phonons (three LO, six TO). At the Γ-Point of the Brillouin zone, the irreducible representation of optical phonon is: Γ= A1+2B1+E1+2E2 [44, 45],whereas E modes are double fold degenerate, namely $E_2$ (high) and $E_2$ (low) related with the motion of oxygen and Zn sub-lattice respectively [44]. $A_1$ and $E_1$ are IR and Raman-active, $E_2$ non-polar branches and only Raman active while $B_1$ modes are silent (inactive mode), i.e. both IR and Raman inactive. Fig.10 represent the room temperature Raman spectra of ZnO for optical properties investigation and the peaks position appeared in all sample in Raman are summarized in Table 3. In Fig.10 indicates seven peaks, sharp and intense peak is detected at 98.6 cm$^{-1}$ (P1) and 438.5 cm$^{-1}$(P6) are assigned to the $E_2^{low}$ (low frequency) and $E_2^{high}$ (high frequency) and are related with the signal of Zn and oxygen sub-lattice respectively. Strong $E_2^{high}$ optical mode indicates a characteristic Raman active peak for the hexagonal wurtzite phase of ZnO and good crystallinity [46]. The small peak due the multi-phonon process is detected at 332.4 cm$^{-1}$(P3), is consigned to $E_2^{high}$-$E_2^{low}$ and recognized as a second order vibration mode ascending from the zone boundary phonon. Additional low intensity peak is perceived at 382.8 cm$^{-1}$ (P4) and allocated to be $A_1$(TO) mode. In additional there is a two small peak around at 411.1 cm$^{-1}$ (P5) and 581.2 cm$^{-1}$ (P7) assigned as $E_1$(TO) and $E_1$(LO) respectively [47, 48]. This $E_1(LO)$ mode is related with the existence of defect corresponding to interstitial Zn, Oxygen vacancies, or their complex and point out that the prepared ZnO nano-rods are of good optical feature. We also observed secondary mode vibration at 150 cm$^{-1}$ and assigned as a $2E_2^{low}$ [49]. There is no obvious change in the Raman peaks except a shift in $E_1(LO)$.

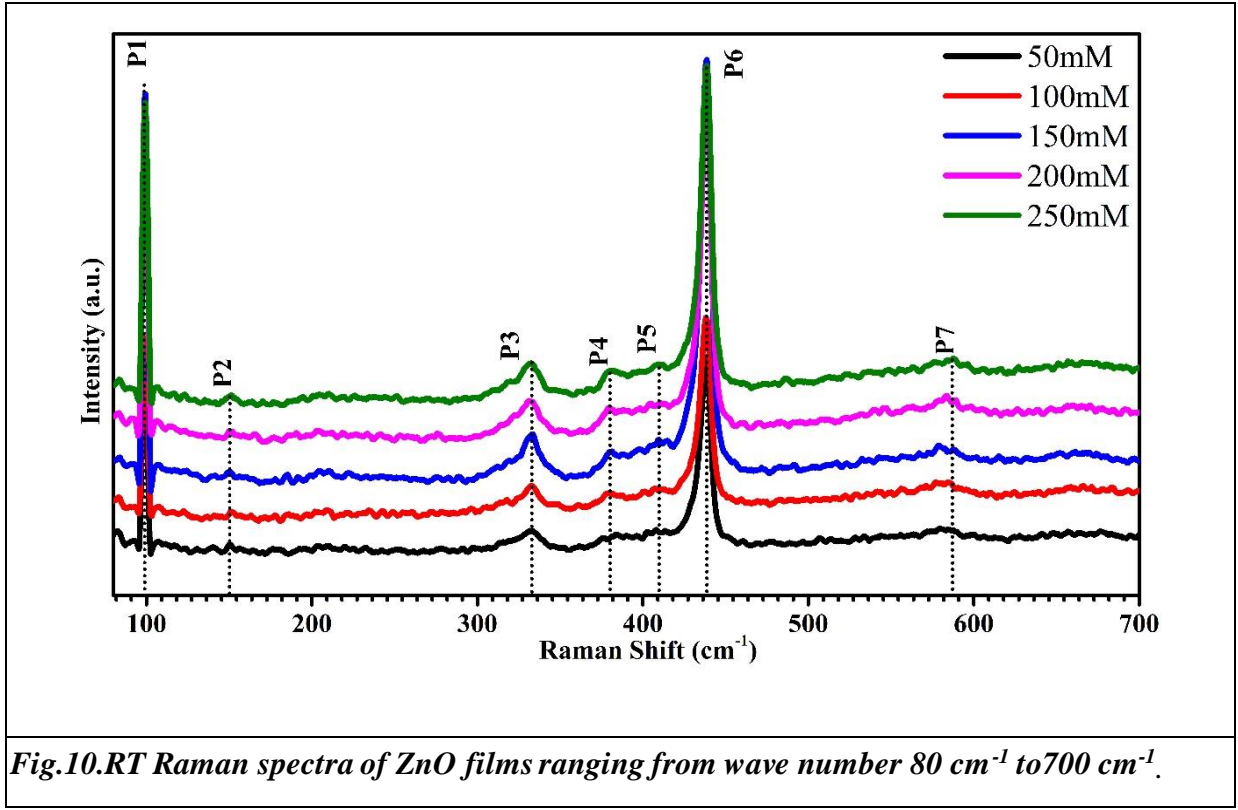

*Fig.10. RT Raman spectra of ZnO films ranging from wave number 80 cm$^{-1}$ to 700 cm$^{-1}$.*

| Peak no. | Positions of the vibrations bands (cm$^{-1}$) | | | | | Assignment |
|---|---|---|---|---|---|---|
| | ZnO (50mM) | ZnO (100mM) | ZnO (150mM) | ZnO (200mM) | ZnO (250mM) | |
| 1 | 98.64 | 99.17 | 99.37 | 99.37 | 99.37 | $E_2$(low) |
| 2 | 150.38 | 151.62 | 150.38 | 151.62 | 151.62 | $2E_2$(low) |
| 3 | 332.49 | 333.59 | 332.59 | 333.44 | 333.59 | $E_2$(high)-$E_2$(low) |
| 4 | 382.81 | 380.42 | 380.97 | 380.97 | 382.21 | $A_1$(TO) |
| 5 | 411.28 | 411.28 | 411.32 | 412.52 | 411.97 | $E_1$(TO) |
| 6 | 438.57 | 438.28 | 438.57 | 438.42 | 438.71 | $E_2$(high) |
| 7 | 581.26 | 581.26 | 580.02 | 584.84 | 587.32 | $E_1$(LO) |

*Table III. Summarize the Phonon mode frequencies (in units of cm$^{-1}$) of wurtzite ZnO films deposited at different concentrations.*

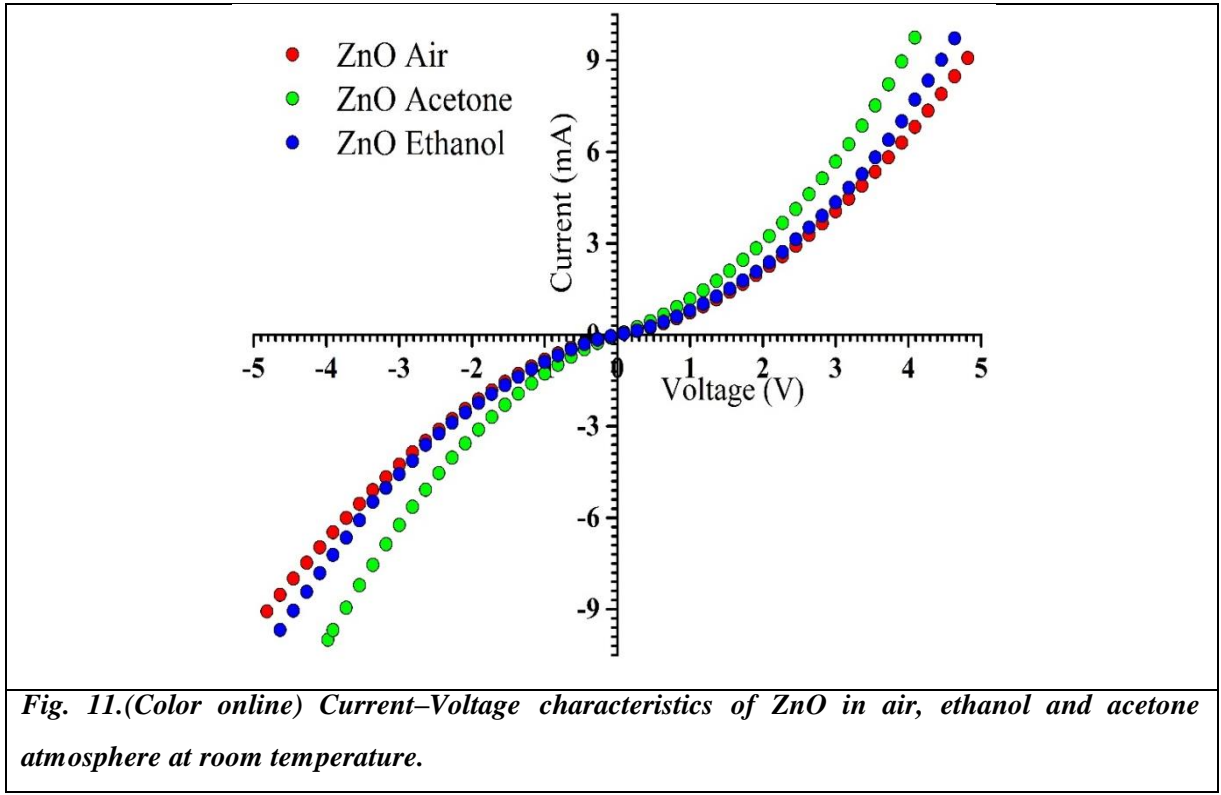

*Fig. 11.(Color online) Current–Voltage characteristics of ZnO in air, ethanol and acetone atmosphere at room temperature.*

Fig.11 depicts the current–voltage (*I–V*) characteristics for ZnO in air, acetone and ethanol atmosphere. The *I–V* features appear to be symmetric with respect to the bias. Consistent and accurate results could only be obtained with a combination of Ohm's law for low voltages, as Ohm's law is generally valid for the low-voltage intrinsic regime of a low-conductivity material [50]. The entire *I-V* characteristics reveal good rectification behaviour of the Schottky diodes. The *fig*.11 shows the room temperature *I-V* characteristic of ZnO nano-rods in different environmental condition i.e. in presence of air, acetone and ethanol. ZnO films shows the nearly linear graph and follows the Ohm's law for low voltage, in all condition.

Metal-Oxide semiconductor is one of the good gas sensors, which work on the principle of the change in electrical conductivity due to chemisorption of the tested gas on the surface of the sensors [51]. In case of *n*-type ZnO based sensors, which belongs to the change in electrical conductivity for detecting the gases. When the ZnO nano-rods expose to air, oxygen molecules absorb on the surface of nano-rods it is because of oxygen deficiency in Zinc oxide nano-rods and form oxygen ions such as $O^{2-}$, $O_2^-$ or $O^-$ ions by capturing the electron from the conduction band, which result in increase of resistance [52]. When ZnO nano-rods expose to other atmosphere i.e. ethanol or acetone atmosphere, these gases adsorbed by the surface of ZnO nano-rods, they capture the oxygen ions from the surface of nano-rods and form a by-product which gives free electron to the conduction band [52, 53]. As a result increase in free electron in conduction band which increases electrical conductivity. Fig.11 shows considerable change in current when the ZnO film expose to ethanol and acetone environment at room temperature. In case of acetone current is more as compare to ethanol, it is because of acetone easily converted into vapors at room temperature and so easily detected by sensor. The current in reverse bias and forward bias is higher in case of acetone and ethanol atmosphere as compared to air, indicate the increase in carrier concentration by changing the atmosphere. Form here we conclude that ZnO nano-rods is one of the good sensor for detection acetone and ethanol at room temperature.

## 4. Conclusion

Simple technique for the well-ordered growth of ZnO nano-needle and nano-rods on glass substrates by controlling merely concentration of ZnO is established. The XRD studies indicates the films are preferentially *c*-axis oriented. Reitveld refinement confirms alternation in the lattice constant with increasing the concentration. The SEM images clearly indicates the change in the ZnO morphology with different concentration with lower deposition temperature (120ºC) and time (90min), needle flowers at (50mM and 100mM) and rods-flowers at (150mM to 250mM). Depending on the morphologies, which is one of the reason to effect the optical properties, naturally band-gap was found to vary. Lower the concentration higher the bandgap and vice versa. PL data indicates the intensities and defect levels changes with concentration. The presence of defects formed in extremely non-stability conditions had a substantial influence on the luminescence of ZnO. We verified a superior control only not on the morphology however also on the defect levels for nano-rods by controlling concentration. Such sample defect level alterations will significantly assistance the applications of ZnO nano-needles/rods in light emission, opto-electronic devices, biological classification, display devices, *etc*. I-V characteristics vary with measurement environment, indicating the potential sensor applications.

**Acknowledgments**

This work was supported by the Department of Science and Technology, India by awarding the prestigious 'Ramanujan Fellowship' (SR/S2/RJN-121/2012) to the PMS. PMS is thankful to Prof. Pradeep Mathur, Director, IIT Indore, for encouraging the research and providing the necessary facilities. We are thankful to Prof. Vasant Sathe, IUC-DAE Consortium for Scientific Research, Indore for his help to do Raman measurement of the samples. We are also thankful to Dr. T. Pradeep (DST Nanoscience unit) for TEM measurements. The help received for PL measurement from Dr. Vipul Singh is also acknowledged.